\documentclass[conference]{IEEEtran}
\usepackage[utf8]{inputenc}
\usepackage[T1]{fontenc}
\usepackage{microtype}
\usepackage{graphicx}
\usepackage{dblfloatfix}
\usepackage{filecontents}
\usepackage[noadjust]{cite}
\usepackage[hidelinks, bookmarks=false, draft]{hyperref}
\usepackage[keeplastbox]{flushend}
\usepackage{amsmath}
\usepackage{url}
\usepackage{multirow}
\usepackage[table,xcdraw]{xcolor}
\usepackage{textcomp}

\begin{document}

\title{Reframing Societal Discourse \\as Requirements Negotiation: \\ Vision 
Statement}

\author{\IEEEauthorblockN{Kurt Schneider\IEEEauthorrefmark{1}, Oliver 
Karras\IEEEauthorrefmark{1}, Anne Finger\IEEEauthorrefmark{2}, and Barbara 
Zibell\IEEEauthorrefmark{2}}
	\IEEEauthorblockA{\IEEEauthorrefmark{1}Software Engineering Group, Leibniz Universität Hannover, 30167 Hannover, Germany\\
		Email: \{kurt.schneider, oliver.karras\}@inf.uni-hannover.de}
	\IEEEauthorblockA{\IEEEauthorrefmark{2} Department of Planning and 
	Sociology of Architecture, Leibniz Universität Hannover, 30167 Hannover, 
	Germany\\ Email: \{a.finger, b.zibell\}@igt-arch.uni-hannover.de}}

\maketitle

\begin{abstract}
Challenges in spatial planning include adjusting settlement patterns to 
increasing or shrinking populations; it also includes organizing food delivery 
in rural and peripheral environments. Discourse typically starts with an open 
problem and the search for a holistic and innovative solution. Software will 
often be needed to implement the innovation. Spatial planning problems are 
characterized by large and heterogeneous groups of stakeholders, such as 
municipalities, companies, interest groups, citizens, women and men, young 
people and children. Current techniques for participation are slow, laborious 
and costly, and they tend to miss out on many stakeholders or interest groups.

We propose a triple shift in perspective: (1) Discourse is reframed as a 
requirements process with the explicit goal to state software, hardware, and 
organizational requirements. (2) Due to the above-mentioned characteristics of 
spatial planning problems, we suggest using techniques of requirements 
engineering (RE) and CrowdRE for getting stakeholders (e.g. user groups) 
involved. (3) We propose video as a medium for communicating problems, 
solution alternatives, and arguments effectively within a mixed crowd of 
officials, citizens, children and elderly people.

Although few spatial planning problems can be solved by software alone, this 
new perspective helps to focus discussions anyway. RE techniques can assist in 
finding common ground despite the heterogeneous group of stakeholders, e.g. 
citizens. Digital requirements and video are well-suited for facilitating 
distribution, feedback, and discourse via the internet. In this paper, we 
propose this new perspective as a timely opportunity for the spatial planning 
domain -- and as an increasingly important application domain of CrowdRE.

\end{abstract}


\IEEEpeerreviewmaketitle

\section{Negotiation in Public Spatial Planning}
Public decision-making often takes place in the tension between innovative, 
competing ideas, and controversial opinions. Informal interest groups tend to 
be large and heterogeneous, with implications for the possible modes of 
discourse: There are commercial and other private stakeholders representing 
different roles, ages, positions, and intentions. Complex dependencies and 
technical language will not reach all of them. In many spatial planning 
problems, it is not easy to involve all relevant citizens and get all affected 
groups involved in the process of discourse and decision-making. Only a small 
selection of stakeholders participates in decision-making processes. Typically, 
only a few active ``speaker'' comment and express their opinions in interviews, 
focus groups, or feedback. Due to the varying educational backgrounds, 
stakeholders including informal interest groups need to be supported in 
different ways. This will enable them to contribute to informed decisions. 

Town hall or other plenary meetings are costly, difficult to schedule and thus, 
rare; a discourse that relies on continuing communication and discussion will 
suffer from long pauses, oblivion in-between meetings, and the need to spend a 
significant part of the precious meeting time on updates and explanations. 
Larger stakeholder groups will hardly ever meet.

When citizens and members of different societal groups meet, there is not 
necessarily a common language or common ground to build on. Providing 
information in an adequate form is essential to bridge those gaps. This 
information will have to convey basic facts; stakeholders need to be informed 
about different alternative solutions, both in the large and in detail. 

This situation has a lot in common with the communication gap between 
customers, requirements engineers, and developers of software 
\cite{Jackson.1995}. However, the style of interaction between software 
customers and developers tends to be more solution-oriented and less 
emotionally loaded than spatial planning discourses. In a software project, 
parties start from the assumption that they want to create useful and usable 
software. The overall goal will be similar across different proposals.

Therefore, we suggest \textit{pretending to plan for software that will support 
solving the spatial planning problem}. In order to specify that software, 
diverging opinions on solution alternatives must be resolved, too. Although 
software may be only a small part of the solution, the stronger focus on 
technical services instead of human convictions can help to make the discourse, 
in general, more effective. We are aware of the limitations of this proposal; 
technocratic approaches are definitely not the best or only solutions to 
societal problems. However, the attitude of defining and discussing and 
negotiating the supposedly easy technical part could indeed stimulate and 
facilitate discourse. In fact, many of the traditional, long-lasting discourses 
lead to a second round of defining software support. When that happens, 
building on existing requirements \textit{already adopted during the first 
round of discourse} can save time and money. We approach this vision 
in an interdisciplinary way.

RE is a discipline with a long history of taking its application domains very 
seriously. It provides cross-cutting techniques for all phases of the  
requirements analysis. Interviews and workshops are already used by both 
requirements engineers and by spatial planners; other techniques have not 
gained much attention in the planning field: Examples include goal modeling, 
exploration of work processes and user interfaces through use cases or mockups. 
We propose to consider the entire range of formal and informal RE techniques, 
and videos in particular. Videos are the best documentation option for 
communication between people who are globally distributed \cite{Ambler.2002}. 
This medium provides the benefit of capturing extensive verbal and nonverbal 
information \cite{Karras.2016b}. Furthermore, videos are easy to share and can 
be used by anyone \cite{Karras.2017}.

In order to motivate our approach, we introduce spatial planning in 
Sec. \ref{ch:2} and show a real example in slightly more detail (Sec. 
\ref{ch:3}). In Sec. \ref{ch:4}, we identify relevant RE techniques. In 
particular, we highlight techniques from CrowdRE which closely match the needs 
of public discourse. Sec. \ref{ch:5} presents related work. In Sec. \ref{ch:6}, 
we outline our plans for evaluation, which will require long-term 
interdisciplinary collaboration. Sec. \ref{ch:7} concludes the paper.

\section{The Domain of Public Spatial Planning}
\label{ch:2}
\begin{figure}[!b]
	\centering
	\includegraphics[width=1.0\linewidth]{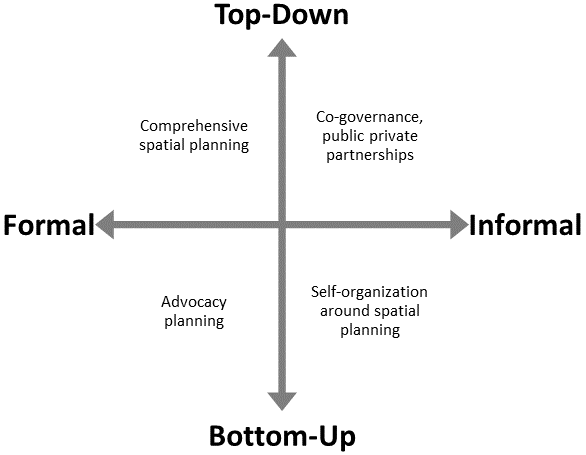}\hfill
	\caption{Different types of spatial planning, based on \cite{Horelli.2016}}
	\label{figure:1}
\end{figure}

Spatial Planning takes place in a field of different initiators, performers, 
and dimensions of formalization. Planning processes can be initiated by the 
state or municipalities (top-down) as well as by citizens or other private 
stakeholders (bottom-up). Formal processes are characterized by binding and 
formalized procedures and outcomes, while informal processes are not regulated 
by law and their outcomes can be binding only for the administration, but not 
for other stakeholders (see \figurename{ \ref{figure:1}}). Due to political 
decision processes, participation of public agencies, public participation, 
regulated sequences and weighing alternatives, planning processes often take 
years from creating the project or plan idea to realization 
(\textit{challenge 1}).

In bottom-up processes, communication and coordination of the different 
stakeholders - mostly non-decision makers - is quite important. In top-down 
spatial planning, the role of public participation differs in formal and 
informal processes. According to Arnstein \cite{Arnstein.1969}, formal planning 
processes include participatory elements on the level of informing and 
consultation, which means information on the plan is provided; public 
agencies, authorities, and citizens are entitled to formulate their views 
during the plan preparation procedure. However, informal approaches are 
comparatively free of legal limitations and open for the active shaping of the 
project. Public participation concentrates on achieving consensus 
\cite{Pahl.2008}.

\begin{figure}[htbp]
	\centering
	\includegraphics[width=1.0\linewidth]{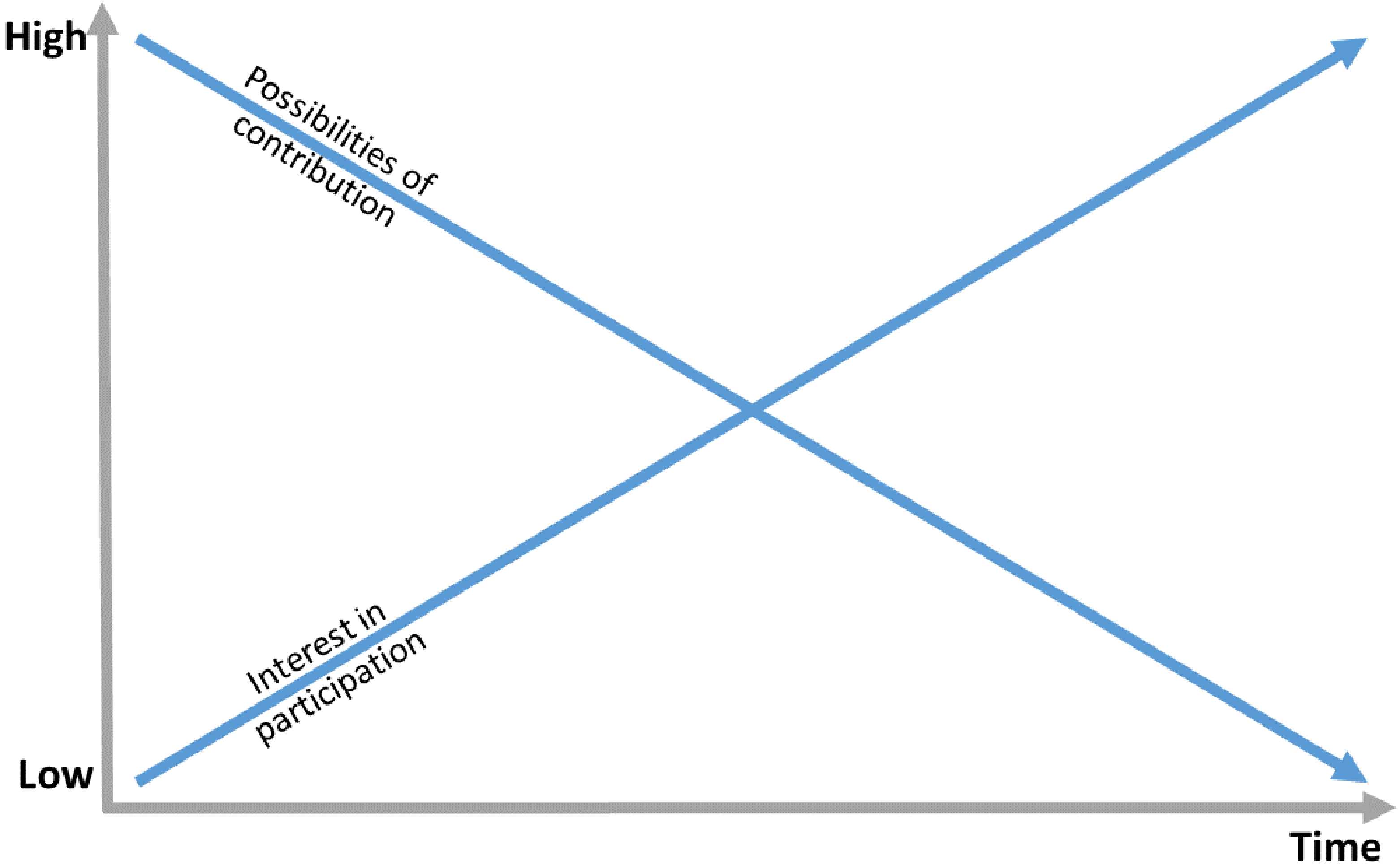}\hfill
	\caption{Paradox of participation, based on \cite{Reinert.1997}}
	\label{figure:2}
\end{figure}

Although public interest in projects and participation is comparatively low at 
the beginning of the planning processes, the possibility for citizens to yield 
their interests is very high. The possibility of influencing the project will 
drop during the project period, but the public interest will rise (see 
\figurename{ \ref{figure:2}}); this is called the paradox of participation 
\cite{Reinert.1997}. Therefore, early public participation makes sense and 
could be highly effective, but struggles with attracting participants 
(\textit{challenge 2}).

In comparison to Requirement Engineering, public formal planning as a process 
is less ``user-oriented'' and less oriented towards stakeholder participation. 
Instead, it focuses on the general welfare and weighing of private and public 
interests. As the request for public participation is getting stronger during 
the last years, many municipalities implemented informal participation in 
planning processes. CrowdRE with its participative elements offers potential to 
be implemented in participation phases of informal and bottom-up spatial 
planning processes. The combination with Requirement Engineering seems to be 
even more useful if the project's realization can benefit by the Requirement 
Engineering techniques for eliciting, negotiating, and validating varying 
perspectives. 
The following example is used to demonstrate where CrowdRE could be added to 
spatial planning projects.

\section{Real Example: Food Supply in Rural Environments}
\label{ch:3}
Many rural areas in Germany are facing severe challenges regarding the local 
food supply. Due to demographic and structural changes, retailers are moving 
from smaller to bigger villages, to small towns, or even concentrate on 
greenfield developments outside of settlements. Consequently, local residents 
in rural areas face long-distance driving to stores and supply facilities. As a 
consequence, residents become more dependent on motorized private or public 
transport. Public transportation in a sparsely populated area is often 
insufficient.

\subsection{Proposed Solution of the Example Planning Problem}
In this setting, local citizens cooperating with transdisciplinary research 
processes develop ideas of a new local supply infrastructure. The following 
example ``LieferBar'' \cite{Jagenteufel.2013} aims at finding out whether and 
to which extent social networks as well as the readiness to cooperate among 
stakeholders – in view of difficult demographic and economic preconditions – 
may contribute to context specific solutions for local food supply in affected 
municipalities.

The modular infrastructure of ``LieferBar'' is based on a vending machine 
concept, providing drawers for different goods adjusted to local and current 
needs. Producer, distributors and service providers – such as bakeries, 
farmers, grocers, postal services, pharmacies, libraries or bicycle shops – 
rent a drawer to offer their goods and services. The approach leads to lower 
personnel cost, and local residents have shorter ways. The supply 
infrastructure can combine constant supply, as for imperishable goods or 
permanent services, with on-call purchase. Coordination and on-demand supply 
require a software application to coordinate supply and demand, and potentially 
optimize delivery routes and schedules.

\subsection{Challenge of Participatory Decision-Making -- In a Crowd}
In situations like rural supply on demand, there are several options available: 
(a) Goods could be delivered by different stores independently and according to 
a statistic of past demands; alternatively, (b) customers could order specific 
demands individually over the internet; stores could coordinate their 
deliveries. (c) For urgent needs, even private parties and citizens could offer 
to pick up goods in town if they happen to be there anyway. All of these 
options have advantages and disadvantages for various stakeholders. Developing 
a solution in the traditional way would take years; ideas that were innovative 
in the beginning, such as ``LieferBar'' may be outdated by the time of 
implementation. After the principal decision has been taken, software 
development starts, almost anew.

In this paper, we propose to use RE and crowd-based approaches for speeding up 
the process and for stimulating a lively early debate. In particular, we want 
to include the issue of software support in the early phases of planning 
alternatives. In the above-mentioned example, options (a,b,c) will require 
substantially different software support. By focusing on the software 
requirements early reverses the typical order. We expect a more concrete 
discussion much earlier in the process and an excellent interface to a 
potential software acquisition or development phase.

The intention behind the software-inspired process is to offer rich 
information earlier and in a more captivating way, thus getting stakeholders 
involved, discussing on a more concrete level. Ideally, the presentation of 
potential solutions via mockups, videos, and simulations should help to avoid 
the participation paradox (see \figurename{ \ref{figure:2}}). Furthermore, 
crowd-based approaches, which have been successfully established in planning 
processes can be used to make the evaluation of realized projects much more 
attractive.

\section{Relevant RE Techniques}
\label{ch:4}
In this section, we briefly review a selection of well-known RE techniques with 
respect to their suitability for spatial planning. Other techniques may be 
considered as well, but are not listed due to space restrictions. \tablename{ 
\ref{table:1}} summarizes the focus of following RE techniques and indicates 
their relevance for spatial planning.

\subsection{Established RE Techniques}
\textbf{Interviews and workshops} are used in both spatial planning and RE. 
Those techniques are best suited when a defined and limited set of people 
should be asked intensely about their requirements and opinions. Interviews 
tend to address one or only a few participants representing the same 
stakeholder or interest group. In RE, interviews are typically used in 
elicitation and validation. Workshops offer an opportunity to stimulate 
discussion among several people or different stakeholder groups. Workshops are 
typically used in creative decisions or in requirements negotiation, where 
diverging opinions are allowed to clash and be discussed. The success of a 
workshop crucially depends on preparation. For example, preparing and 
discussing goal models assists stakeholders in focusing.

\textbf{Goal models} show dependencies between stakeholders due to resources, 
tasks, goals, and soft goals. There is more than one notation for goal models 
(e.g., i* \cite{Yu.1997} and KAOS \cite{Darimont.1997}), but the principle use 
of goal models is similar. Since interest groups have goals by definition, 
capturing and modeling those goals may be a good idea and stimulating for 
certain phases of discussion. However, goal model notations are not 
straight-forward to understand for non-technical stakeholders, and goal models 
do not easily scale up for large problems with many stakeholders or interest 
groups. This example of a popular RE technique shows the limitations of 
adopting RE techniques for spatial planning: Representations must be 
comprehensible to the citizens and stakeholders they are supposed to support. 

\textbf{A persona} is a real-looking profile description of a (mostly 
fictitious) person. It should be easy to understand. Thus, personas are a good 
medium for overcoming abstract explanations or discussions that are difficult 
to follow for many people. Creating concrete, but fictitious representatives of 
difficult-to-grasp and often abstract entities may be a useful technique for 
spatial planning. This technique enables stakeholders and developers to develop 
empathy and deeper emotional understanding for the situation and demands of the 
modeled person or entity such as residents or local suppliers.

\begin{table*}[!h]
	\centering
	\caption{Relevance of RE Techniques for Spatial Planning}
	\label{table:1}
	\begin{tabular}{|c|l|l|}
		\hline
		\rowcolor[HTML]{DFDFDF} 
		\textbf{\begin{tabular}[c]{@{}c@{}}RE technique\end{tabular}} & 
		\multicolumn{1}{c|}{\cellcolor[HTML]{DFDFDF}\textbf{Purpose}} & 
		\multicolumn{1}{c|}{\cellcolor[HTML]{DFDFDF}\textbf{Relevance for 
				spatial planning}} \\ \hline
		Mockup & \begin{tabular}[c]{@{}l@{}}Rough, static sketch of GUI or 
			other \\ perceivable aspect\end{tabular} & 
		\begin{tabular}[c]{@{}l@{}}Any device or user interface stakeholders 
			may face in one of the discussed options. \\ This can include paper 
			forms and technical interfaces.\end{tabular} \\ \hline
		Prototype & \begin{tabular}[c]{@{}l@{}}Executable software with limited 
			and \\ focused functionality or user interface\end{tabular} & 
		Applicable in later phases, requires some level of programming. For 
		complex tasks. \\ \hline
		\begin{tabular}[c]{@{}c@{}}Vision Video\end{tabular} & 
		\begin{tabular}[c]{@{}l@{}}Represents envisioned scenarios,\\ 
			alternative options, or narrows in on \\ aspects under 
			discussion\end{tabular} & \begin{tabular}[c]{@{}l@{}}Appropriate 
			for a 
			broad audience; very concrete. Can be produced at different cost \\ 
			levels (Smart­phone to professional).\end{tabular} \\ \hline
		Interview & \begin{tabular}[c]{@{}l@{}}Focused transfer of information 
			from \\ interviewee to interviewer\end{tabular} & Essential for 
		in-depth elicitation; does not scale up to larger crowds. \\ \hline
		Workshop & \begin{tabular}[c]{@{}l@{}}Exchanging opinions, using group 
			\\ dynamics for developing new \\ proposals\end{tabular} & 
		\begin{tabular}[c]{@{}l@{}}Size determines the character and potential 
			outcome. Participation of crucial \\ stakeholders is essential; 
			should 
			not take too long.\end{tabular} \\ \hline
		\begin{tabular}[c]{@{}c@{}}Goal Models\end{tabular} & 
		\begin{tabular}[c]{@{}l@{}}Mainly for eliciting and negotiating \\ 
			goals and rationale\end{tabular} & Applicable to small expert 
		subgroups, not to a crowd of all stakeholders. \\ \hline
		Use Case & \begin{tabular}[c]{@{}l@{}}Semi-formal representation of \\ 
			interaction between actors and system\end{tabular} & 
		\begin{tabular}[c]{@{}l@{}}Several variants available; can be selected 
			and tailored to different situations of use. \\ Not as concrete as 
			videos, but can be more detailed.\end{tabular} \\ \hline
	\end{tabular}
\end{table*}

\textbf{Use cases} are a very popular style of describing an interaction 
between an actor and a system for the purpose of fulfilling a goal of that 
actor \cite{Cockburn.2000}. Since use cases are mainly textual and written in 
natural language which is only slightly regulated, most people would be able to 
understand the core of a use case when it is explained by a developer. However, 
there are more aspects to a ``full-fledged'' \cite{Cockburn.2000} use case, 
such as trigger, guarantee, pre-condition etc.. Those aspects are more 
sophisticated to understand. For that reason, use cases are either reserved for 
an advanced group of stakeholders -- or use cases must be adapted, simplified, 
or complemented by more accessible representations. There are a large number of 
alternative representations of scenarios \cite{Alexander.2005}. They are more 
or less adequate to bridge the communication gap between the heterogeneous 
crowds of stakeholders in spatial planning.

\textbf{Videos} are a comparatively new addition to the toolbox of scenario 
representations. They are intended to represent visions, alternative options, 
or usage of a future system in a very concrete way. Fricker et al. 
\cite{Fricker.2015} propose videos of stakeholder discussions as an ideal 
medium for improved requirements communication. Karras et al. 
\cite{Karras.2016b} emphasize the potential of vision videos which are used as 
early prototypes that require no coding at all. Xu et al. \cite{Xu.2012} 
proposed an evolving video artifact in which portions of mockup videos are 
replaced by screencast video clips step by step. Many authors who advocate 
videos for RE highlight their ability to communicate effectively with various 
different people. Thus, videos are useful to provide concrete contextual 
information in town hall meeting as well as for the crowd.

\textbf{Mockups and prototypes} are useful in general. They can assist in 
different phases of interest group discussions. Videos are a new type of 
prototypes; they require no programming and show a selected scenario in a 
dynamic representation \cite{Karras.2017}.

\subsection{Techniques of CrowdRE}
Many RE techniques were initially tailored towards individual software 
development for a defined customer. In product development, market-driven 
development, and in spatial planning, however, there are various solution 
proposals available, and rather sizable crowds of stakeholders who should be 
enabled to participate. Since the definition of \textit{CrowdRE}, several 
techniques were identified that are specifically appropriate for a large and 
mixed crowd of stakeholders. Spatial planning problems usually have exactly 
that profile.

CrowdRE starts from the \textit{assumption} that there is a crowd of 
participants who are able and willing to communicate via electronic media. 
Thus, they can receive electronic messages at short notice and have the 
technical infrastructure for responding. A crowd could emerge from a generic 
social network, e.g. Facebook. It could as well be organized via a designated 
tool or platform on the internet. This basic assumption facilitates the 
following four services, according to Groen and Koch \cite{Groen.}, see 
\tablename{ \ref{table:2}}:

\begin{table}[htbp]
	\centering
	\caption{CrowdRE-Services and Examples in Spatial Planning}
	\label{table:2}
	\begin{tabular}{|l|l|}
		\hline
		\rowcolor[HTML]{DFDFDF} 
		\multicolumn{1}{|c|}{\cellcolor[HTML]{DFDFDF}\textbf{Service} and 
		examples} & 
		\multicolumn{1}{c|}{\cellcolor[HTML]{DFDFDF}\textbf{Purpose} and 
		examples} \\ \hline
		\begin{tabular}[c]{@{}l@{}}\textbf{Crowdsourcing}\\ Web-based 
		prototyping, \\ 
		usability testing, online \\ discussion, forums, \\ community 
		management, \\ design decisions\end{tabular} & 
		\begin{tabular}[c]{@{}l@{}}\textbf{Using the crowd as input}\\ Shaping 
		idea, 
		public participation, \\ weighting alternatives, searching \\ for 
		practical applications of \\ proposed solution\end{tabular} \\ \hline
		\begin{tabular}[c]{@{}l@{}}\textbf{Text Mining}\\ Social network 
		analysis, \\ 
		recommender systems \\ (requirements writing, \\ prioritization, 
		voting)\end{tabular} & \begin{tabular}[c]{@{}l@{}}\textbf{Extracting 
		higher-level concepts} \\ \textbf{from social media}\\ Situational 
		analysis, 
		goals, \\ weighting alternatives, opinion \\ forecasts\end{tabular} \\ 
		\hline
		\begin{tabular}[c]{@{}l@{}}\textbf{Usage Mining}\\ Data mining of use 
		or \\
		queries, prototyping, \\vision videos\end{tabular} & 
		\begin{tabular}[c]{@{}l@{}}\textbf{Monitoring user behavior and} \\ 
		\textbf{drawing conclusions}\\ Public display, public decisions, \\ 
		public 
		participation\end{tabular} \\ \hline
		\begin{tabular}[c]{@{}l@{}}\textbf{Motivational Instruments}\\ Feeback, 
		gamification\end{tabular} & 
		\begin{tabular}[c]{@{}l@{}}\textbf{Attracting 
		stakeholders}\\ Participation alternatives, incentives \\for 
		participation\end{tabular} \\ \hline
	\end{tabular}
\end{table}

These services can substantially improve the spatial planning process: 
Instruments used for motivation can attract interest groups and individuals at 
an earlier phase, mitigating the above-mentioned paradox of participation (see 
\figurename{ \ref{figure:2}}). Once involved, stakeholders have the opportunity 
to receive information in various formats and respond (see 
\textit{assumption}). The above-mentioned techniques of RE can also be applied 
and combined. The activated crowd can now receive any electronic document, 
video, or message that seems appropriate. Their ability to respond 
instantaneously adds an important driver to the discussion. Speed becomes 
important in order to sustain the interchange within the attention span of the 
public. 

Exchange on the internet facilitates asynchronous communication, overcoming 
the challenge of finding a suitable time and place to meet in person. 
Stakeholders can view visualizations and provide their feedback and opinions 
without attending a meeting. Software requirements arise as a side-effect of 
supporting the decision-making process, which gives software development a 
clear head start.

Not all stakeholders will have the infrastructure, the permission, or the 
desire to interact in the crowd. Therefore, RE and CrowdRE techniques are not 
supposed to replace face-to-face communication, but to improve preparation, 
information, and evolution of alternative options, since the turnaround time 
can be shrunk. There will still be town hall meetings; they can benefit by 
using the material emerging from crowd collaboration, thus making better use of 
the invested time and effort. Videos, text mining of discussions, monitoring of 
data and usage can tie together virtual and face-to-face discourse.

\section{Related Work}
\label{ch:5}
We suggest supporting crowds of stakeholders in spatial planning problems by 
treating them like a crowd of software customers, even if no software is 
supposed to be developed. 

Arias et al. \cite{Arias.1999, Arias.1998} approached a similar problem 
from the opposite angle: Citizens of Boulder discussing flood mitigation 
options started with a highly concrete ``language of pieces'' (trees, houses 
etc.) to visualize their ideas. As the discussion went on, they accepted more 
abstract visualizations (colored blocks, blue lines for Boulder Creek), which 
could easily be manipulated and analyzed on a computer. In this paper, we argue 
for an approach that does not assume long-term collaborative learning but 
offers concrete visualizations like personas, mockups, and videos. They can 
help to bridge the communication gap caused by an abstract and vague 
presentation.
Brill et al. \cite{Brill.2010} had compared video-based requirements to text 
and use case-based requirements and found both useful, but with complementary 
strengths and weaknesses. In our current student software projects, a Cyber 
Crime Unit of the Hannover police was involved. A different project dealt with 
the local hospital Radiology Unit, yet another one with the North-West German 
Volleyball League. Understanding these different domains turned out to be much 
easier using 2-minute vision videos that were created after the third week of 
requirements elicitation. Video seems to be well-suited for focusing 
interactions on complex future software. 
Koch et al. \cite{Koch.2016} present a very inspiring case of using personas 
to a planning situation very similar to our food supply example. They designed 
a strategy for volunteers to pick up food for their neighbor. Our current work 
also starts with traditional RE techniques but then proposes to explicitly 
extend it to CrowdRE and videos. We also propose to adopt a software-oriented 
perspective from the beginning, no matter whether a software is already on the 
plan or not.
Evans-Crowley and Hollander \cite{Evans.2010} described the possibilities of 
working with virtual web-based participation in spatial planning processes. The 
authors recommend planners to embrace new digital tools but also point out that 
the access to high-speed web and devices or the acceptance of digital 
approaches leads to different accessibilities for citizens to digital 
participation processes.

\section{Findings and Evaluation Plan}
\label{ch:6}
\subsection{Summarizing the Core Contribution}
Spatial planning deals with large crowds of stakeholders who should participate 
in decisions that will shape their environments. The planning process extends 
over months or years, and by the time stakeholders get aware of the 
alternatives, their influence has already decreased (see 
\figurename{ \ref{figure:2}}). Spatial planning wants to create and support 
highly innovative and timely improvements, but there are many challenges 
preventing the effective participation of interest groups and individuals.

RE, on the other hand, is a discipline focusing on the exchange with domain 
experts, customers, and many different stakeholders -- for the purpose of 
developing software that matches user needs and requirements. CrowdRE even 
offers techniques and services tailored to make use of the internet, and 
improve communication drastically.

We envision applying RE and CrowdRE techniques to spatial planning processes 
even if it is not yet decided to buy or develop software for any of the new and 
innovative proposals. We think that this turn in perspective will make 
discussions more concrete (less vague and abstract) and more attractive to 
citizens. Since many innovative solutions finally \textit{do} require software 
support, this approach also provides faster and better-prepared entrance 
into the software development phase.

\subsection{Status of the Vision and Planned Evaluation}
This work is part of the 4.5M€ \textit{Mobilise} initiative of 
the Universities of Hannover and Braunschweig in Germany. In the \textit{Mobile 
Man} project (part of that initiative), five faculties collaborate in order to 
investigate intelligent mobility at the intersection of computer science, 
RE, geo-informatics, law, ethics, and spatial planning. 

We saw a surprising tendency of discussing societal issues from a purely 
technical perspective, e.g.: \textit{What can autonomous driving do? Can I 
predict from your past travel behavior where you will be going tomorrow?} 
Despite this technology-driven debate, we are convinced that the fast-growing 
opportunities from autonomous driving, intelligent and individualized 
navigation, mobile information and booking systems will soon call for the input 
from empowered citizens and stakeholders. It is only fair and economic to use 
known and established techniques (of RE and CrowdRE) for getting stakeholders 
involved in organizing more effective participation.

We plan to apply CrowdRE techniques and services in several situations during 
the \textit{Mobile Man} project. Spatial planning will be an early application 
domain. Since planning processes are very long-lasting, we do not have them at 
our disposal. Instead, we will apply the proposed reframing to a discourse 
\textit{among participating scientists} first. They can be considered a 
technology-friendly selection of stakeholders for a pretest. Once this first 
hurdle will be taken, the authors of this paper (requirements engineers and 
planners) will apply the approach when the planners get called into the next 
applicable situation.

\begin{figure}[htbp]
	\centering
	\includegraphics[width=1.0\linewidth]{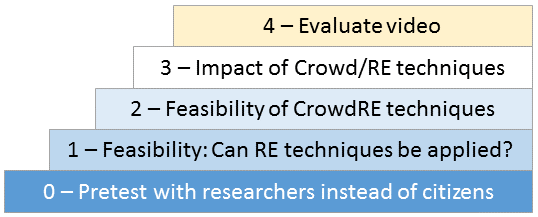}\hfill
	\caption{Planned steps of evaluation}
	\label{figure:3}
\end{figure}

Evaluation is planned to go through several progressive steps, as sketched in 
\figurename{ \ref{figure:3}}. After the above-mentioned pretest with 
researchers, we will investigate (1) whether the RE techniques can be applied 
in a way allowing the general public to participate effectively. 
(2) CrowdRE techniques are more sophisticated and require access to computers 
or smartphone, and an ability and willingness to engage in a process of 
informed decision-making and technology-supported feedback. (3) The impact of 
all techniques will be assessed qualitatively and finally quantitatively. If 
possible, we will use the new techniques as treatment of an experiment and 
compare its influence with reference to a control group which uses traditional 
planning and participation techniques. This observational approach will be 
triangulated by a survey in both groups, soliciting opinions on the techniques 
used. (4) As a final component, we will ask for contents and analyze whether 
the treatment group can recall and explain complex information better when it 
is presented as videos rather than text.

\subsection{Threats to Validity Ahead: Research Challenge in the Crowd}
There are numerous potential threats to validity that require our attention in 
this research. We are aware evaluation will take a lot of time and effort. 
Therefore, we have prioritized evaluation activities and will try to take one 
step at a time (see \figurename{ \ref{figure:3}}). Field studies in real 
planning situations can be treated like case studies and analyzed according to 
the guidelines provided by Runeson et al. \cite{Runeson.2012}.

\section{Conclusion}
\label{ch:7}
RE helps spatial planning in decision-making by offering tools and techniques. 
To handle large crowds of stakeholders, digital artifacts are particularly 
welcome: Artifacts can be displayed in meeting rooms, sent to homes and 
smartphones. Feedback can easily be given and channeled back to moderators. 
Mechanisms of CrowdRE can be applied and developed further. The result of this 
interdisciplinary approach always includes software requirements; they can be 
used directly for developing innovative software, which will speed up the 
implementation of a decision. In other cases, requirements remain a by-product 
and annotation of the original planning decision. We are convinced that this 
can empower citizens in their new role as software customers and, thus, focus 
societal discussions.

\section*{Acknowledgment}
This work was carried out within the interdisciplinary \textit{Mobile Man} 
project at Leibniz Universität Hannover and was supported by the 
German Research Foundation (DFG) project \textit{ViViReq}.

\bibliographystyle{IEEEtran}
\bibliography{IEEEabrv,ref}

\begin{thebibliography}{10}
\providecommand{\url}[1]{#1}
\csname url@samestyle\endcsname
\providecommand{\newblock}{\relax}
\providecommand{\bibinfo}[2]{#2}
\providecommand{\BIBentrySTDinterwordspacing}{\spaceskip=0pt\relax}
\providecommand{\BIBentryALTinterwordstretchfactor}{4}
\providecommand{\BIBentryALTinterwordspacing}{\spaceskip=\fontdimen2\font plus
\BIBentryALTinterwordstretchfactor\fontdimen3\font minus
  \fontdimen4\font\relax}
\providecommand{\BIBforeignlanguage}[2]{{%
\expandafter\ifx\csname l@#1\endcsname\relax
\typeout{** WARNING: IEEEtran.bst: No hyphenation pattern has been}%
\typeout{** loaded for the language `#1'. Using the pattern for}%
\typeout{** the default language instead.}%
\else
\language=\csname l@#1\endcsname
\fi
#2}}
\providecommand{\BIBdecl}{\relax}
\BIBdecl

\bibitem{Jackson.1995}
M.~Jackson, ``The world and the machine,'' in \emph{Proceedings of the 17th
  International Conference on Software Engineering}.\hskip 1em plus 0.5em minus
  0.4em\relax ACM, 1995.

\bibitem{Ambler.2002}
S.~W. Ambler, \emph{{Agile Modeling: Effective Practices for eXtreme
  Programming and the Unified Process}}.\hskip 1em plus 0.5em minus 0.4em\relax
  New York: Wiley, 2002.

\bibitem{Karras.2016b}
O.~Karras, S.~Kiesling, and K.~Schneider, ``{Supporting Requirements
  Elicitation by Tool-Supported Video Analysis},'' in \emph{24th IEEE
  International Requirements Engineering Conference}.\hskip 1em plus 0.5em
  minus 0.4em\relax IEEE, 2016.

\bibitem{Karras.2017}
O.~Karras, C.~Unger-Windeler, L.~Glauer, and K.~Schneider, ``{Video as a
  By-Product of Digital Prototyping: Capturing the Dynamic Aspect of
  Interaction},'' in \emph{3rd International Workshop on Usability and
  Accessibility Focused Requirements Engineering}.\hskip 1em plus 0.5em minus
  0.4em\relax IEEE, 2017.

\bibitem{Horelli.2016}
L.~Horelli, ``{Engendering Urban Planning in Different Contexts -– Successes,
  Constraints and Consequences},'' in \emph{5th Engendering International
  Conference: Engendering HABITAT III Facing the Global Challenges in Cities,
  Climate Change and Transport}, 2016.

\bibitem{Arnstein.1969}
S.~R. Arnstein, ``{A Ladder of Citizen Participation},'' \emph{Journal of the
  American Institute of Planners}, vol.~35, no.~4, 1969.

\bibitem{Pahl.2008}
E.~Pahl-Weber, D.~Henckel \emph{et~al.}, \emph{{The Planning System and
  Planning Terms in Germany: A Glossary}}.\hskip 1em plus 0.5em minus
  0.4em\relax Verl. d. ARL, 2008, vol.~7.

\bibitem{Reinert.1997}
A.~Reinert and H.~Sinning, ``{Mobilizing Civil Competence. The ÜSTRA Citizens'
  Panel on Public Transport in Hanover (in German: Mobilisierung der Kompetenz
  der Bürgerinnen und Bürger. Das Bürgergutachten ÜSTRA zum öffentlichen
  Nahverkehr in Hannover)},'' 1997.

\bibitem{Jagenteufel.2013}
S.~Jagenteufel and C.~Sandherm, ``{Future-Compliant Regional Development in
  lower Saxony (in German: Wettbewerbsbeitrag zum ALR-Hochschulpreis 2013
  ``zukunftsfähige land- und regionalentwicklung in niedersachsen'')},'' in
  \emph{{Competition Entry for ALR University Award}}, 2013.

\bibitem{Yu.1997}
E.~S. Yu, ``{Towards Modelling and Reasoning Support for Early-Phase
  Requirements Engineering},'' in \emph{Proceedings of the 3rd IEEE
  International Symposium on Requirements Engineering}.\hskip 1em plus 0.5em
  minus 0.4em\relax IEEE, 1997.

\bibitem{Darimont.1997}
R.~Darimont, E.~Delor, P.~Massonet, and A.~van Lamsweerde, ``{GRAIL/KAOS: An
  Environment for Goal-Driven Requirements Engineering},'' in \emph{Proceedings
  of the 19th International Conference on Software Engineering}.\hskip 1em plus
  0.5em minus 0.4em\relax ACM, 1997.

\bibitem{Cockburn.2000}
C.~Alistair, \emph{{Writing Effective Use Cases}}.\hskip 1em plus 0.5em minus
  0.4em\relax Addison-Wesley, 2000.

\bibitem{Alexander.2005}
I.~F. Alexander and N.~Maiden, \emph{{Scenarios, Stories, Use Cases: Through
  the Systems Development Life-Cycle}}.\hskip 1em plus 0.5em minus 0.4em\relax
  John Wiley \& Sons, 2005.

\bibitem{Fricker.2015}
S.~A. Fricker, K.~Schneider, F.~Fotrousi, and C.~Thuemmler, ``{Workshop Videos
  for Requirements Communication},'' \emph{Requirements Engineering}, 2015.

\bibitem{Xu.2012}
H.~Xu, O.~Creighton, N.~Boulila, and B.~Bruegge, ``{From Pixels to Bytes:
  Evolutionary Scenraio Based Design with Video},'' in \emph{ACM SIGSOFT 20th
  International Symposium}, 2012.

\bibitem{Groen.}
E.~C. Groen and M.~Koch, ``{How Requirements Engineering can benefit from
  crowds},'' \emph{Requirements Engineering Magazine}, 2016.

\bibitem{Arias.1999}
E.~G. Arias, H.~Eden, G.~Fischer, A.~Gorman, and E.~Scharff, ``{Beyond Access:
  Informed Participation and Empowerment},'' in \emph{{Proceedings of the 1999
  Conference on Computer Support for Collaborative Learning}}.\hskip 1em plus
  0.5em minus 0.4em\relax International Society of the Learning Sciences, 1999.

\bibitem{Arias.1998}
E.~Arias, K.~Schneider, and S.~Thies, ``{A Continuum Approach: From Language of
  Pieces to Virtual Stakeholders},'' in \emph{World Conference on Artificial
  Intelligence in Education}, 1998.

\bibitem{Brill.2010}
O.~Brill, K.~Schneider, and E.~Knauss, ``{Videos vs. Use Cases: Can Videos
  Capture More Requirements under Time Pressure?}'' in \emph{Requirements
  Engineering: Foundation for Software Quality}.\hskip 1em plus 0.5em minus
  0.4em\relax {Springer Berlin Heidelberg}, 2010, vol. 6182.

\bibitem{Koch.2016}
M.~Koch, S.~He{\ss}, A.~He{\ss}, and D.~P. Magin, ``{Digital Innovations of
  citizens for citizens -- Design Thinking or Citizen Science? (in German:
  Digitale Innovationen von B{\"u}rgern f{\"u}r B{\"u}rger -- Design Thinking
  oder Citizen Science?)},'' \emph{UP 2016}, 2016.

\bibitem{Evans.2010}
J.~Evans-Cowley and J.~Hollander, ``{The New Generation of Public
  Participation: Internet-Based participation Tools},'' \emph{Planning Practice
  \& Research}, vol.~25, no.~3, 2010.

\bibitem{Runeson.2012}
P.~Runeson, M.~Host, A.~Rainer, and B.~Regnell, \emph{{Case Study Research in
  Software Engineering: Guidelines and Examples}}.\hskip 1em plus 0.5em minus
  0.4em\relax John Wiley \& Sons, 2012.

\end{thebibliography}

\end{document}